\documentclass[floatfix,
 aps, 
 amsmath,amssymb,
 reprint, 
 citeautoscript
]{revtex4-1}
\usepackage[utf8]{inputenc}
\usepackage{graphicx}
\usepackage{float}
\usepackage{enumitem}
\usepackage{bm}
\usepackage{bbold}
\usepackage{cases}
\usepackage{upgreek}
\usepackage{xcolor}
\usepackage{siunitx}

\usepackage[normalem]{ulem}

\usepackage[mathscr]{euscript}
\usepackage{algorithm}
\usepackage{algpseudocode} 

\usepackage{textgreek}

\makeatletter

\def\citation#1{\@for\@tmpkey:=#1\do{%
  \expandafter\gdef\csname mainref@\@tmpkey\endcsname{}}}
\def\methodsmarker{\gdef\citation##1{}}
\newcommand\MethodsCitations{%
  \if@filesw\immediate\write\@auxout{\string\methodsmarker}\fi}
\newif\ifbib@mainlist \bib@mainlisttrue
\let\bib@orig@lbibitem\@lbibitem
\long\def\bib@skipitem#1\BibitemShut#2{}
\def\@lbibitem[#1]#2{%
  \ifbib@mainlist
    \@ifundefined{mainref@#2}{\expandafter\bib@skipitem}
                             {\bib@orig@lbibitem[#1]{#2}}%
  \else
    \@ifundefined{mainref@#2}{\bib@orig@lbibitem[#1]{#2}}
      {\global\advance\c@NAT@ctr\@ne\expandafter\bib@skipitem}%
  \fi}
\newcommand\methodsbibliography{%
  \bib@mainlistfalse
  \begingroup\let\label\@gobble\@input@{\jobname.bbl}\endgroup}
  
\def\@email#1#2{%
 \endgroup
 \patchcmd{\titleblock@produce}
  {\frontmatter@RRAPformat}
  {\frontmatter@RRAPformat{\produce@RRAP{*#1\href{mailto:#2}{#2}}}\frontmatter@RRAPformat}
  {}{}
}%
\makeatother

\usepackage{textgreek}

\makeatletter
\def\@email#1#2{%
 \endgroup
 \patchcmd{\titleblock@produce}
  {\frontmatter@RRAPformat}
  {\frontmatter@RRAPformat{\produce@RRAP{*#1\href{mailto:#2}{#2}}}\frontmatter@RRAPformat}
  {}{}
}%
\makeatother

\usepackage{hyperref}

\hypersetup{
  colorlinks   = true, 
  urlcolor     = magenta, 
  linkcolor    = magenta, 
  citecolor   = purple 
}

\begin{document}

\preprint{AIP/123-QED}

\title{Memory-gated transient anomalous transport of quasiparticles out of equilibrium}

\author{Pranay Venkatesh}
\affiliation{Department of Chemistry, University of Colorado Boulder, Boulder, Colorado}

\author{Tianchu Li}
\affiliation{Department of Chemistry, University of Colorado Boulder, Boulder, Colorado}

\author{Srijan Bhattacharyya}
\affiliation{Department of Chemistry, University of Colorado Boulder, Boulder, Colorado}

\author{Andr\'es Montoya-Castillo}
\email{Andres.MontoyaCastillo@colorado.edu}
\affiliation{Department of Chemistry, University of Colorado Boulder, Boulder, Colorado}
\date{\today}

\begin{abstract}
Anomalous transport is ubiquitous across condensed matter, materials science, and biophysics. Yet, in far-from-equilibrium settings, what determines when, how much, and in which direction transport departs from diffusion remains a fundamental open question. By adopting the dispersive Holstein model of small polarons as representative of a wide class of dissipative systems, we show that a signed bath memory time governs the direction and the strength of the transient departure, and that when it vanishes, diffusion emerges without an asymptotic delay. Tuning the dissipative environment from resonant to off-resonant with the carrier density of states, transport turns from subdiffusion to superdiffusion as the memory time changes sign. Microscopically, we trace this memory effect to a competition between polaron formation and coherent polaron motion. These findings establish a unified memory-based description of transient anomalous transport, offering a path to control nonequilibrium quasiparticle flow through environmental tuning.
\end{abstract}

\maketitle



Understanding and controlling the transport of charge, energy, and mass through a medium is a grand challenge across materials design~\cite{noriega2013general, tsurumi2017coexistence, franchini2021polarons}, condensed matter physics~\cite{gross2017quantum, peng2023exploiting, gopalakrishnan2024non}, and the study of glassy and biological systems~\cite{debenedetti2001supercooled, berthier2023modern}. It underlies technologies ranging from optoelectronic devices~\cite{chin2013role, zhao2025stable} to neuromorphic computing and information processing~\cite{markovic2020physics, Williams2024}. Transport can be normal or anomalous. With a mean-squared displacement that grows linearly in time, normal (diffusive) transport is characterized by a time-independent diffusion coefficient; anomalous transport departs from this behavior. Carriers may move subdiffusively, trapped and released by disorder or strong coupling to their environment, or superdiffusively, propagating coherently over extended times and distances. The consequences are profound: subdiffusion limits how far excitations travel through energy materials before they recombine, and disorder can arrest transport altogether~\cite{Anderson1958, altman2018many}. Yet, the various stages of far-from- to near-equilibrium transport after a nonequilibrium quench, the extent of their universality, and their underlying mechanisms remain largely unknown. 

A carrier injected far from equilibrium need not relax to thermal equilibrium at all~\cite{Srednicki1994} and integrability, localization, and many-body scars can each prevent it~\cite{Anderson1958, Rigol2007}. When relaxation does occur, it culminates in a \textit{hydrodynamic limit}, in which a steady-state transport law emerges. For systems whose hydrodynamic limit is diffusion, the emergence of diffusion additionally serves as a proxy for thermalization~\cite{landau1981statistical}. Even for systems that do reach the hydrodynamic limit, nonequilibrium relaxation can occur over long timescales, as in prethermal plateaus~\cite{Gring2012}, Griffiths bottlenecks~\cite{Agarwal2015}, or dipole-constrained dynamics~\cite{gromov2024fracton}. What sets the direction and duration of such transient departures from diffusion is generally unclear. Answering this requires knowing both how strongly a carrier exchanges energy with its environment and how the environment's response to the carrier's motion shapes its approach to the hydrodynamic limit.

Small polarons---carriers dressed by a local distortion of their host lattice---dominate energy and charge transport in transition metal oxides, organic crystals, and conjugated polymers~\cite{franchini2021polarons}. Since they form and move as a result of excitations exchanging energy with the lattice, they offer a natural setting in which to track a carrier from its initial nonequilibrium injection to its hydrodynamic limit. Advances in ultrafast spectroscopy are only now starting to make this passage directly observable, by measuring the mean-squared displacement of photogenerated carriers with nanometer precision~\cite{ginsberg2020spatially, penwell2017resolving}. Recent experiments reveal that phonon engineering can tune materials so that nonequilibrium carrier transport after photoexcitation is either slower~\cite{carneiro2017excitation} or faster~\cite{liu2021nonequilibrium, zhang2022ultrafast} than their long-time diffusive speed. Stimulated by these advances, theory has begun to interrogate the origin of the transient departure from diffusion~\cite{bhattacharyya2025nonequilibrium, mendes2026coherent}. Little is known about what controls the direction of this departure, superdiffusive in some materials and subdiffusive in others, or how far it extends. A microscopic theory could thus unlock the potential to systematically control transient transport over device-relevant length scales~\cite{ginsberg2020spatially, liu2021nonequilibrium} in addition to observing it.

\begin{figure*}[htb!]
    \centering
    \includegraphics[width=\linewidth]{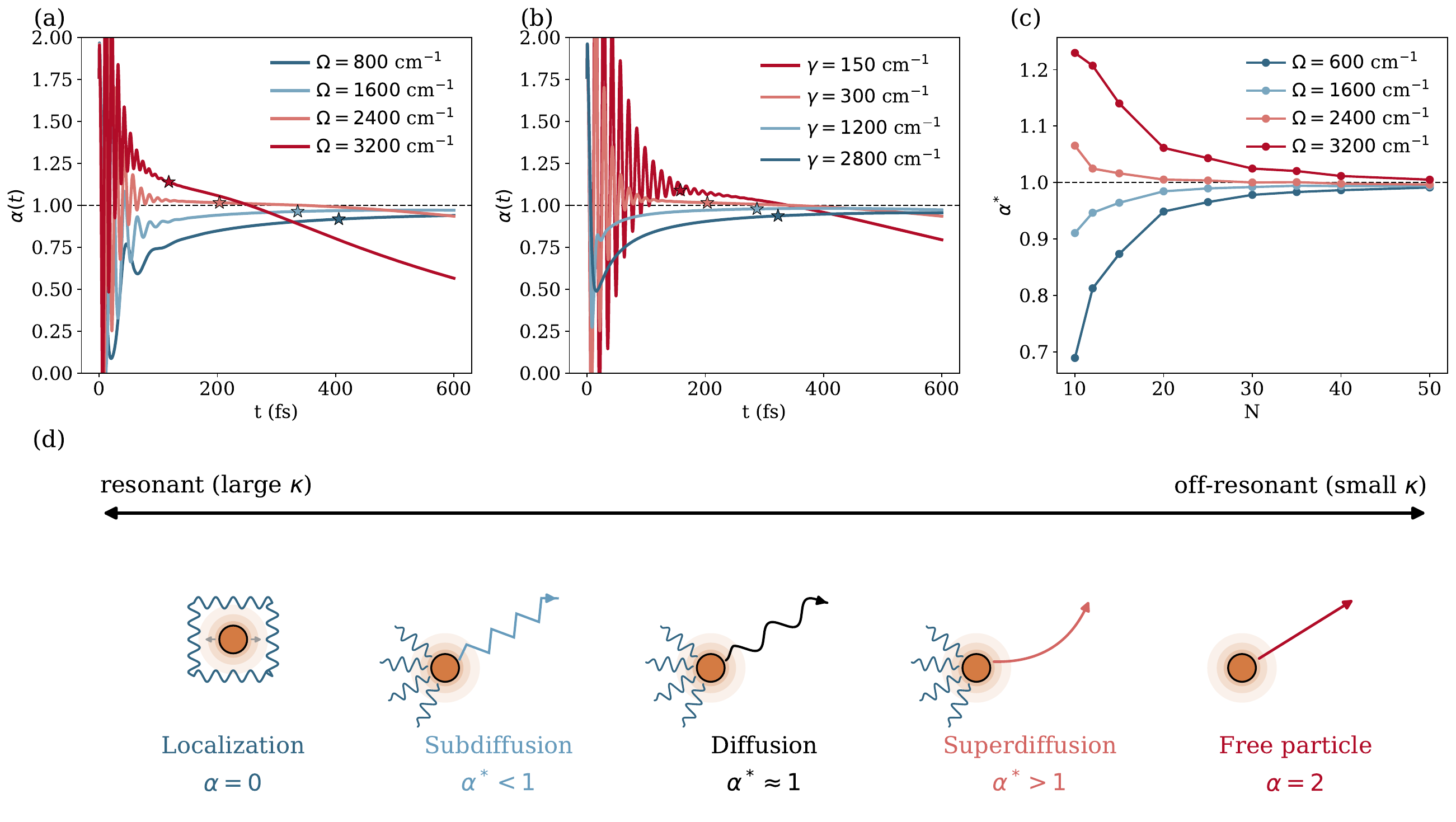}
    \caption{\textbf{Tuning the phonon bath off resonance carries transient transport from subdiffusion to superdiffusion.} \textbf{a},~Time-dependent transport exponent $\alpha(t)$ for increasing phonon frequency $\Omega$ at fixed damping $\gamma = 300~\unit{cm}^{-1}$. Stars mark $\alpha^*$, the value of $\alpha(t)$ at the onset of finite-size effects. \textbf{b},~$\alpha(t)$ for increasing $\gamma$ at fixed $\Omega = 2400~\unit{cm}^{-1}$. Broadening the mode returns transport from the superdiffusive to subdiffusive regime. \textbf{c},~$\alpha^*$ versus lattice size $N$ at $\gamma = 300~\unit{cm}^{-1}$ for varying $\Omega$. On both sides of the crossover, $\alpha^*$ approaches the diffusive value $\alpha^* = 1$ (dashed line) as $N$ grows. \textbf{d},~Schematic of transport regimes ordered by the resonance $\kappa$ between the phonon bath and the carrier, from Anderson localization ($\alpha = 0$) in the resonant limit to free-particle motion ($\alpha = 2$) in the off-resonant limit. \textbf{a},\textbf{b},~$N = 15$. Other parameters as in Methods.}
    \label{fig:fig1}
\end{figure*}

Here, we adopt the dispersive Holstein model of small polarons as a controlled incarnation of a carrier relaxing from its initial excitation and transitioning to the hydrodynamic limit. We identify which factors control the direction and extent of its transient departure from diffusion and find that tuning the phonon bath from resonant to off-resonant with the carrier tunes transport continuously from asymptotically long-lasting transient subdiffusion to superdiffusion, with Anderson localization and ballistic motion as the two limits. Mapping the lattice size over which this departure persists across bath parameters, we find a crossover seam between these two regimes where the particle thermalizes to diffusion swiftly. We show that a single bath memory time $\tau_M$, derived from the memory kernel of the carrier's dynamics, predicts the sign and magnitude of the departure from diffusion and the location of the seam. The sign of $\tau_M$ reflects two competing sources of memory, polaron formation after nonequilibrium carrier injection and the coherent motion of the dressed carrier over several sites. Between these effects lies a crossover where both effects cancel. Because this picture rests only on how a carrier's environment retains memory of its motion, we expect it to apply generally to excitations relaxing in dissipative environments.

\section*{Holstein polarons}

The Holstein model describes how a carrier interacts with a local optical phonon mode~\cite{Holstein1959} and is the workhorse for interrogating small polaron physics~\cite{franchini2021polarons}. The Hamiltonian describes an electronic lattice with $N$ sites,
\begin{equation}
    H = \sum_i \varepsilon_i \hat{c}_i^{\dagger} \hat{c}_i + \sum_{i,j} v_{i,j} \left(\hat{c}_i^{\dagger} \hat{c}_j + \hat{c}_j^{\dagger} \hat{c}_i \right) + \sum_i [H_i^{ph} + H_i^{el-ph}],
    \label{eq:holstein-hamiltonian}
\end{equation}
where each site couples bilinearly to a local phonon, i.e., $H_i^{ph} + H_i^{el-ph} = \Omega_{i} \hat{b}_{i}^{\dagger} \hat{b}_{i} + g_{i} \hat{c}_i^{\dagger} \hat{c}_i (\hat{b}_{i}^{\dagger} + \hat{b}_{i})$. 

While the optical phonons (of frequency $\Omega$) in the Holstein model are infinitely long-lived, real phonons have finite lifetimes~\cite{klemens1966}. Hence, we broaden each mode by connecting it to an auxiliary Markovian bath with damping $\gamma$, thus shortening its lifetime. This maps the single phonon mode per lattice site into a local phonon bath described by a coupling distribution or spectral density where the set of site-dependent coupling constants $\{ g_{i, \zeta}\}$ is described by a Brownian oscillator (Lorentzian) form $J_i(\omega) = \pi\sum_{\zeta} g_{i,\zeta}^2 \delta(\omega - \omega_{i,\zeta}) = \frac{4\lambda\Omega^2 \gamma \omega}{(\omega^2 - \Omega^2)^2 + 4 \gamma^2 \omega^2}$~\cite{garg1985effect, li2024there}.  $\lambda = g^2/\Omega$ represents the reorganization energy of local phonon baths. 

Comparing $J(\omega)$ with the electronic density of states (DOS) allows us to tune the energetic overlap, and thus the resonance conditions between the electron and the phonons by adjusting spectral-density parameters. A phonon of frequency $\Omega$ can scatter the carrier between states only if their energies differ by $\hbar\Omega$, and in the tight-binding band these differences range from $0$ to the bandwidth $4v$. We therefore measure resonance with a single parameter, the thermally weighted overlap between the spectral density and the density of these transition energies $\rho_{el}(\omega)$, $\kappa \equiv \int_0^{\infty} d\omega\ \rho_{el}(\omega) J(\omega) \coth(\frac{\hbar\omega}{2 k_B T})$. $\kappa$ is large when phonon weight lies within the bandwidth and vanishes when it lies above it. For the Brownian oscillator, raising $\Omega$ above $4v$ or narrowing the mode (decreasing $\gamma$) thus pushes the bath off resonance from the carrier. SI.~Sec.~II gives the explicit form of $\rho_{el}$ and maps $\kappa$ over $(\Omega, \gamma)$.

To ground our parameterization of the environment in real materials, we tie the Brownian oscillator bath to the Raman spectra of typical small polaron formers---transition metal oxides and organic semiconductors by varying the pair of parameters $(\gamma, \Omega)$ from $(150, 600)~\unit{cm}^{-1}$ to $(1200, 3200)~\unit{cm}^{-1}$, representing transition metal oxides~\cite{iliev2003role}, perovskites~\cite{leguy2016dynamic} and organic semiconductors~\cite{girlando2010peierls}. Working in the dilute limit of a homogeneous lattice, we set $\varepsilon_i = 0$ and restrict the coupling to a nearest-neighbor constant $v_{ij} = v = 300~\unit{cm}^{-1}$. To specifically probe how the environment controls nonequilibrium relaxation, we quench the system at $t=0$ with the charge localized on a single site and follow its population spread across the lattice with a generalized master equation (GME, see Methods).

The mean-squared displacement (MSD) of a particle directly probes its transport regime. Near equilibrium, one can classify the nature of transport with a time-independent exponent, $\alpha$, that describes the scaling of the MSD with time, ${\rm MSD} \sim t^{\alpha}$, with $\alpha = 2$ indicating ballistic, $1 < \alpha < 2$ superdiffusive, $\alpha = 1$ diffusive, $0 < \alpha < 1$ subdiffusive, and $\alpha = 0$ arrested or localized. Far from equilibrium, this exponent becomes time-dependent, $\alpha \rightarrow \alpha(t)$, and identifies the instantaneous transport regime.

\section*{Nonequilibrium Anomalous Transport}

How does a phonon environment control anomalous transport? To answer this question, we propose a metric that maps how $\alpha(t)$ changes qualitatively with the environment's details. Two features of these dynamics inform our metric: the closest reach to hydrodynamic transport, $\alpha^*$, and its anomalous flavor, superdiffusive or subdiffusive, encoded by ${\rm sgn}(1- \alpha^*)$. $\alpha^*$ is when the population reaches the periodic boundary and the closest an $N$-site lattice comes to hydrodynamic transport, defining the region over which $\alpha(t)$ remains physically meaningful (Methods). $\lvert 1 - \alpha^* \rvert$ measures how far from diffusive transport a carrier remains as it reaches the boundary, and its sign distinguishes subdiffusive (positive) from superdiffusive (negative) transport.

\begin{figure}[htb!]
    \centering
    \includegraphics[width=0.93\linewidth]{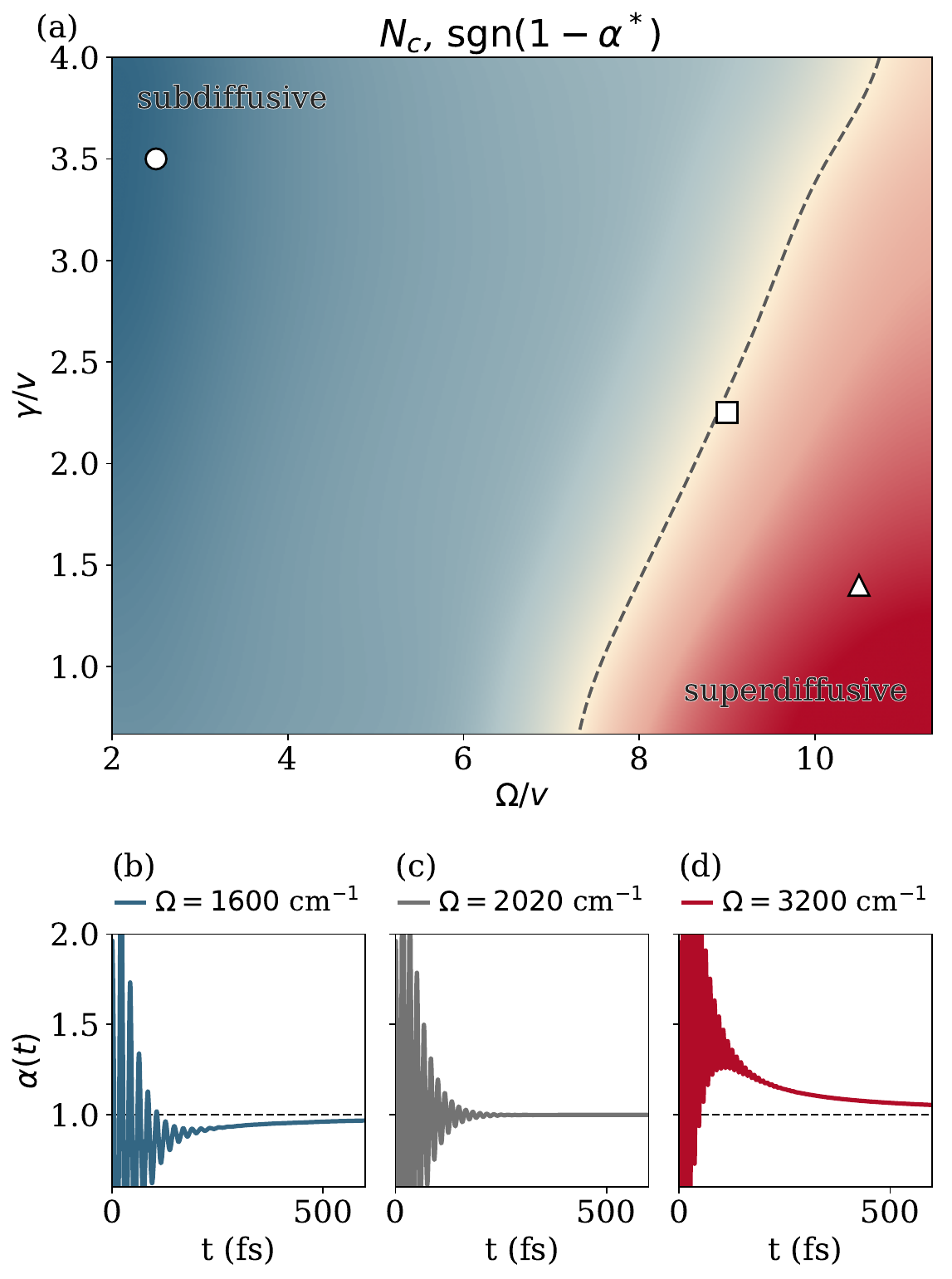}
    \caption{\textbf{Transport diagram for transient anomalous transport.} \textbf{a},~Transport diagram over the plane of ($\Omega$, $\gamma$) in units of hopping $v$. Color indicates the direction of the departure from diffusion (blue: subdiffusive, $\alpha^* < 1$; red: superdiffusive, $\alpha^* > 1$), and shading its reach, with darker colors for larger characteristic size $N_c$. The dashed line marks the diffusive seam where $N_c$ is minimal. Markers indicate representative resonant (circle), seam (square), and off-resonant (triangle) baths. \textbf{b}--\textbf{d},~Time-dependent transport exponent $\alpha(t)$ in the subdiffusive (\textbf{b}; $\Omega = 1600~\unit{cm}^{-1}$), diffusive (\textbf{c}; $\Omega = 2020~\unit{cm}^{-1}$), and superdiffusive (\textbf{d}; $\Omega = 3200~\unit{cm}^{-1}$) regions, at $\gamma = 150~\unit{cm}^{-1}$. Only at the seam does $\alpha(t)$ settle non-asymptotically onto the diffusive value (dashed line) after the early-time oscillations.}
    \label{fig:fig2}
\end{figure}

We begin with a resonant bath, where the phonon frequency and electronic band overlap significantly. Figure~\ref{fig:fig1}(a) shows $\alpha(t)$ for a 15-site lattice coupled to such a bath ($\Omega = 800~\unit{cm}^{-1}$). Upon injection, the carrier spreads ballistically ($\alpha = 2$). As it begins to exchange energy with the lattice, $\alpha(t)$ drops sharply below unity and then rises slowly (asymptotically) back toward diffusion at $\alpha = 1$~\cite{bhattacharyya2024anomalous, bhattacharyya2025nonequilibrium}. When the carrier reaches the lattice boundary, $\alpha^* < 1$, indicating that polarons thermalize toward diffusion asymptotically and subdiffusively after carrier injection in this parameter regime. 

Detuning the phonon frequency $\Omega$ beyond the edge of the electronic band reduces $\kappa$, destroying resonance. Figure~\ref{fig:fig1}(a) shows that going from $\Omega = 800~\unit{cm}^{-1}$ to $1600~\unit{cm}^{-1}$, the subdiffusive departure weakens, with a shallower dip in $\alpha(t)$ after the initial ballistic spread, and a faster recovery towards $\alpha = 1.0$, so that $\alpha^*$ rises from $\sim 0.89$ to $\sim 0.97$. A carrier coupled to a resonant bath is therefore farther from diffusion at the lattice boundary than one coupled to a less resonant bath, despite both having had the same number of sites over which to relax. While one might expect further detuning the bath to bring $\alpha^*$ closer to unity, the flavor of anomalous transport instead flips. At $\Omega = 2400~\unit{cm}^{-1}$, $\alpha(t)$ approaches the boundary from above, with $\alpha^* \approx 1.02$, and at $\Omega = 3200~\unit{cm}^{-1}$ transport is clearly superdiffusive, with $\alpha^* \approx 1.14$. The early-time dynamics also change character. Off resonance, $\alpha(t)$ oscillates strongly over the first few hundred femtoseconds, reflecting energy exchanged back and forth between the carrier and the long-lived phonon mode. Hence, moving the bath off resonance does not merely shorten the transient subdiffusion but moves transport through a \textit{crossover} into superdiffusion. Before interrogating the nature of this crossover in detail, we first identify what factors cause this crossover and the limiting behaviors of transport away from it. 

Building on the intuition of resonance, one might expect the width of the phonon distribution to tune the overlap with electronic timescales. Indeed, we find that where the phonon weight sits relative to the electronic band sets this crossover, so tuning $\Omega$ is not the only way to reach it. Enhancing phonon interactions that decrease the optical phonon's lifetime (increasing $\gamma$) increases weight in the band as effectively as lowering the peak frequency. The resulting transport (Fig.~\ref{fig:fig1}(b)) responds as our overlap picture predicts: at $\Omega = 2400~\unit{cm}^{-1}$, increasing $\gamma$ past $\sim 600~\unit{cm}^{-1}$ causes $\alpha^{*}$ to return into subdiffusive transport. Thus, raising the bath frequency or prolonging its lifetime produces the same crossover.

How does this transient anomalous transport resolve in the thermodynamic limit of $N \rightarrow \infty$? Figure~\ref{fig:fig1}(c) shows that $\alpha^*$ approaches unity with increasing $N$ on both sides of the crossover. For $N = 10$--$50$, this approach is well described by $\lvert 1 - \alpha^* \rvert = \exp(-N/N_c)$, with $N_c$ empirically measuring how far the imprint of a given bath extends before transport becomes hydrodynamic. Together, they condense each realization of $\alpha(t)$ into two numbers that one can follow continuously as the bath is tuned from resonant to off-resonant.

We find that opposite ends of the resonance axis embody drastically different hydrodynamic behavior (Fig.~\ref{fig:fig1}(d)). In the resonant limit, $\Omega \to 0$, $J(\omega) \to 2\pi\lambda\omega \delta(\omega)$, which describes Gaussian static disorder in the site energies~\cite{mukamel1995principles} and matches the widely used Gaussian disorder model of organic semiconductors~\cite{bassler1993charge}. Here, $\kappa$ diverges, the carrier is scattered at every instant, and transport is arrested by Anderson localization immediately after the quench, with $\alpha = 0$~\cite{Anderson1958}. In the off-resonant limit of $\gamma \to 0$ and $\Omega \to \infty$, the phonon mode can no longer exchange energy with the carrier, and $\kappa \to 0$. A Lang-Firsov polaron transformation~\cite{lang1963kinetic, munn1985theory} shows that the model then reduces to a bare tight-binding Hamiltonian (SI Sec.~IV), where a carrier spreads ballistically with $\alpha = 2$. This is the band-like, polaron-free transport that systems such as carbon nanotubes~\cite{frank1998carbon, javey2003ballistic} and graphene~\cite{mayorov2011micrometer} exhibit. In both limits, we expect $N_c$ to diverge, indicating that diffusion never emerges.

Combining these observations in a transport diagram (Fig.~\ref{fig:fig2}(a)), we summarize how the phonon bath modulates the nature and extent of transient anomalous transport across the $(\Omega, \gamma)$ plane, and return to the crossover region. In the resonant corner ($\Omega/v \lesssim 3$, $\gamma/v \gtrsim 3$; circle), transport is strongly subdiffusive and $N_c$ is largest as the bath approaches the static-disorder limit. In the off-resonant corner ($\Omega/v \gtrsim 8$, $\gamma/v \lesssim 2$; triangle), transport is strongly superdiffusive and $N_c$ grows again as the bath approaches the free-particle limit. Most surprisingly, at the crossover seam, $\alpha^* \approx 1$ for all $N$ we simulate. Figure.~\ref{fig:fig2}(c) shows representative $\alpha(t)$, with early-time oscillations that swiftly decay to $1$ with no asymptotic delay, yielding hydrodynamic diffusive behavior well before the onset of finite-size effects. This behavior lies in stark contrast to the asymptotically delayed onset of diffusion that dominates the other regions of the transport diagram.

This swift collapse into hydrodynamic diffusion appears counterintuitive. One might expect the overdamped, resonant corner, where carriers exchange energy with the bath most vigorously, to reach the hydrodynamic limit fastest. It does not. Instead, transient anomalous transport persists over large lattices across the resonant and off-resonant corners, and only swiftly collapses to diffusion at a narrow seam between them. Hence, how resonant a bath is determines the type of transient anomalous transport observed but not how fast it disappears. The speed of this approach must come from a separate quantity. We hypothesize that this quantity is a hidden timescale, the lag with which the bath's dissipative response acts on the carrier's motion. 

\section*{Memory time directs transport}

\begin{figure}[htb!]
    \centering
    \includegraphics[width=\linewidth]{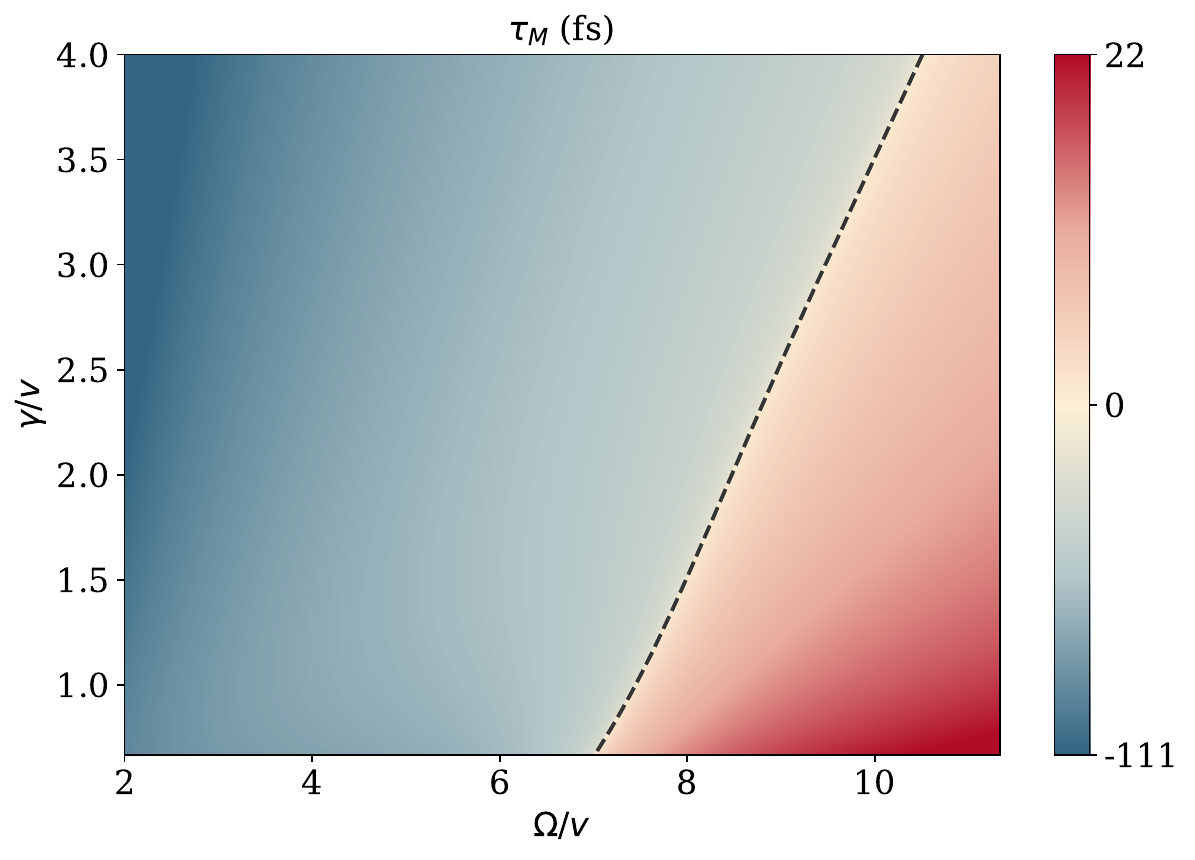}
    \caption{\textbf{Bath memory time reproduces the transport diagram.} Bath memory time $\tau_M$ over the plane ($\Omega$, $\gamma$), in units of the hopping $v$, computed from the two-time perturbative GME memory kernel (Methods). Blue regions indicate $\tau_M < 0$ (subdiffusive approach) and red regions $\tau_M > 0$ (superdiffusive approach). The dashed line marks $\tau_M = 0$, which coincides with the diffusive seam of Fig.~\ref{fig:fig2}. Unlike $N_c$, $\tau_M$ requires no information about lattice size or the threshold that defines $\alpha^*$. Other parameters as in Methods.}
    \label{fig:fig3}
\end{figure}

\begin{figure*}[ht!]
    \centering
    \includegraphics[width=\linewidth]{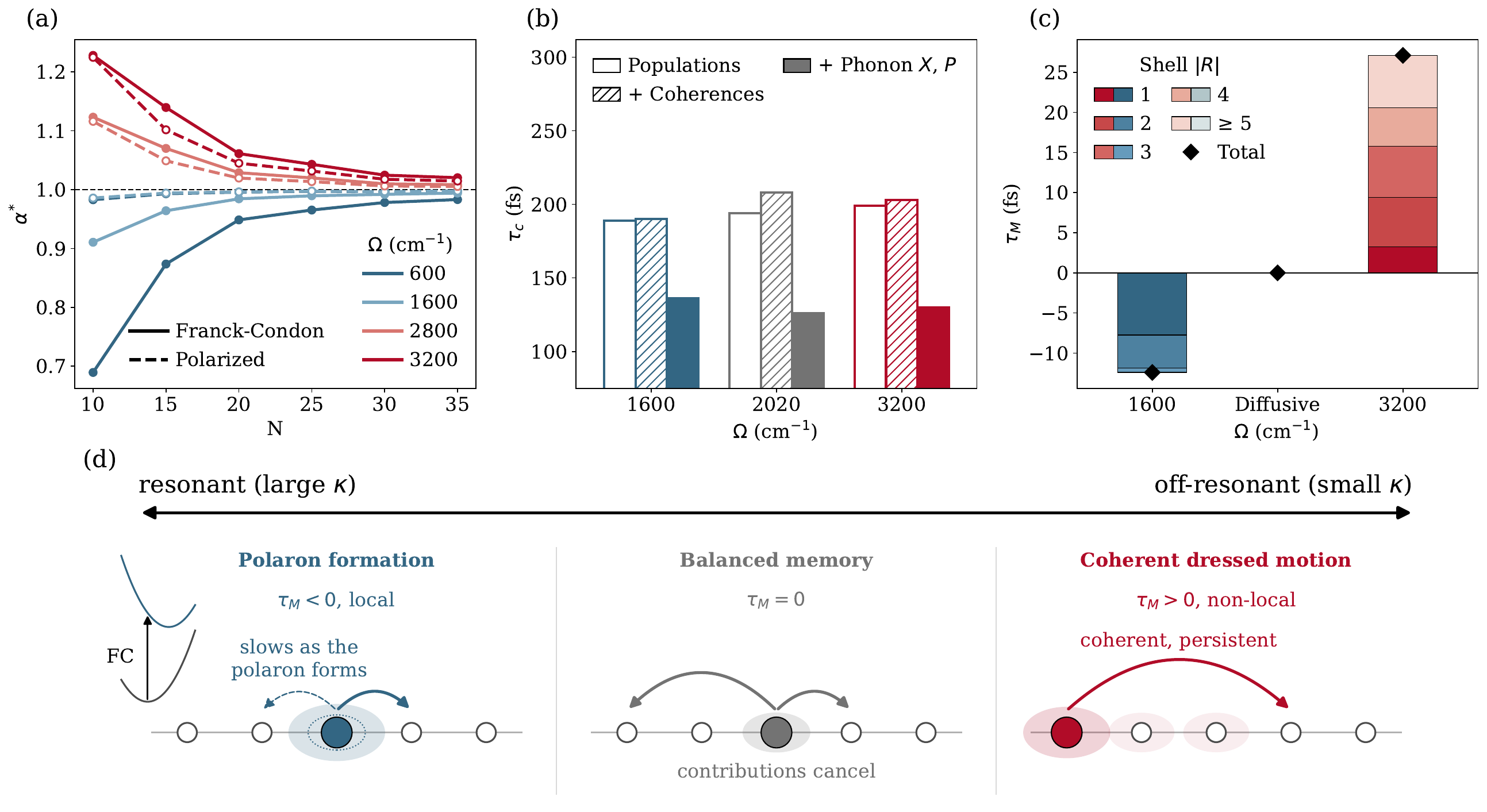} 
    \caption{\textbf{Polaron formation and coherent dressed motion are competing sources of bath memory.} \textbf{a},~$\alpha^*$ versus lattice size $N$ for carriers injected onto an undistorted lattice (Franck-Condon; solid) and onto a lattice polarized around the carrier (dashed). Relaxing the lattice removes the subdiffusive departure but leaves the superdiffusive one largely intact. Dynamics obtained with the population generalized master equation (see Methods), $\gamma = 300~\unit{cm}^{-1}$. \textbf{b},~Lifetime $\tau_c$ of memory kernels obtained by projecting onto the site populations (open), the populations and electronic coherences (hatched), and the populations and local phonon displacement and momentum (filled). $\tau_c$ is the truncation time at which each kernel reproduces the exact MSD to within a fixed tolerance (Methods). \textbf{c},~Decomposition of $\tau_M$ into contributions from kernel elements connecting sites a distance $|R|$ apart (red: positive; blue: negative; darker for smaller $|R|$). Diamonds mark the total. \textbf{d},~Schematic of the two sources of memory. In resonant baths, polaron formation after vertical injection gives the carrier a local head start ($\tau_M < 0$). In off-resonant baths, coherent motion of the dressed carrier over several sites delays its spreading ($\tau_M > 0$). At the diffusive seam the two cancel. Panels \textbf{b} and\textbf{c} use tensor network dynamics (Methods), $\gamma = 150~\unit{cm}^{-1}$, $N = 15$. Other parameters as in Methods.}
    \label{fig:fig4}
\end{figure*}

Such a lag hints at a memory effect. The kernel of the master equation governing the carrier's spread must encode this effect. On a homogeneous lattice, site populations obey
\begin{equation}
  \frac{dP_n(t)}{dt} =  \sum_{m}\int_0^{t} ds\ \mathcal{K}_{n-m}(t-s) P_m(s),
  \label{eq:gme}
\end{equation}
where the memory kernel $\mathcal{K}_{R}(t)$ feeds population from one site into another a distance $R$ away after a delay $t$. The bath influences transport entirely through $\mathcal{K}_R$. It encodes the MSD, $d {\rm MSD}/dt = 2D(t) = \int_0^t dt'\ \tilde{\mathcal{K}}(t')$, where $\tilde{\mathcal{K}}(t) = a^2\sum_R R^2\mathcal{K}_R(t)$ is the distance-weighted kernel. For diffusion, the MSD grows at a constant rate and $\tilde{\mathcal{K}}(t)\propto\delta(t)$, so its Laplace transform is $\hat{\tilde{\mathcal{K}}}(s) = 2D_\infty$. However, a bath with memory saturates the integral only after a delay, adding a first-order correction, $\hat{\tilde{\mathcal{K}}}(s) = 2D_\infty(1 - \tau_M s + \cdots)$. We define this leading non-Markovian correction as the bath memory time $\tau_M = \frac{\int_0^\infty t\tilde{\mathcal{K}}(t)dt}{\int_0^\infty \tilde{\mathcal{K}}(t)dt}$ (Methods). Integrating once more, $\mathrm{MSD}(t) \to 2D_\infty(t-\tau_M)$, revealing that $\tau_M$ is the lag of the MSD behind diffusion, positive if the carrier trails it and negative if it leads. At long times, $\alpha(t) \simeq 1 + \tau_M/t$ relaxes as $1/t$, with $\tau_M$ setting the sign and magnitude of the departure from diffusion.

This analysis goes beyond polaron physics to reveal a universal asymptotic scaling of relaxation to diffusion requiring only the influence of a dissipative environment. This relaxation toward hydrodynamic transport evaporates when the first moment of the memory disappears, leading to a crossover. This crossover and its mechanism survive changes in reorganization energy and temperature, and different bath representations. We further find that these also hold for magnon and mass transport (SI Sec.~V). We therefore expect these to govern other excitations relaxing in structured environments, from excitons~\cite{penwell2017resolving, pandya2021exciton} and polaritons~\cite{xu2023ultrafast, li2026molecular} to the crowded interior of cells, where similar memory effects reshape biochemical kinetics~\cite{ayaz2021non, banani2017biomolecular}.

One might therefore expect $\tau_M$ to locate the crossover for polaron relaxation. It does: our $\tau_M$ map in Fig.~\ref{fig:fig3} faithfully recapitulates that of $1 - \alpha^*$ in Fig.~\ref{fig:fig2}. Remarkably, this agreement required no information about lattice size or the threshold defining $\alpha^*$. $\tau_M$ is negative throughout the resonant region (low $\Omega$, large $\gamma$) and positive in the off-resonant corner (high $\Omega$, small $\gamma$), matching ${\rm sgn}(1-\alpha^*)$ everywhere, with a magnitude that grows toward both corners like $N_c$. The locus $\tau_M = 0$ marks the same diffusive seam ($N_c \to 0$), so resonant baths carry negative memory and off-resonant baths positive, meeting where $\tau_M$ changes sign. Yet, $\tau_M$ locates the seam without explaining it. 

\section*{Competing memory effects}

Given the importance of the bath, we hypothesize that the memory time $\tau_M$ is vibrationally gated, set by how long the phonons take to respond to the carrier. A natural mechanism is environmental dressing of the carrier, the phonon bath in this case. Our nonequilibrium initial condition has so far been a Franck-Condon injection, after which the bath dresses the carrier, an effect the kernel records. Because we have focused on strong polaron formers, fully polarizing the environment around a static carrier would constitute the opposite nonequilibrium initial condition, one that should require less time to relax to an equilibrated but mobile polaron~\cite{kloss2019multiset, mendes2026coherent}. We test this hypothesis (Methods). Yet, Fig.~\ref{fig:fig4}(a) shows that the two sides of the crossover respond oppositely. On the subdiffusive side, the departure vanishes almost entirely---at $\Omega = 600~\unit{cm}^{-1}$ and $N = 10$, $\alpha^*$ rises from $0.69$ to $\sim0.98$ and stays near $1.0$ at every lattice size. This confirms that vibrational relaxation in polaron formation gates negative memory. On the superdiffusive side, however, $\alpha^*$ is largely insensitive to the initial condition, so a second source of memory survives the polarized initial condition (SI~Sec.~III).

Could the memory in the superdiffusive side arise from non-vibrational sources? To test this, we exploit the fact that accounting for all slow variables minimizes the memory kernel lifetime, $\tau_c$~\cite{grabert2006projection, montoya2016a}. Specifically, we test if including electronic coherences minimizes memory lifetimes (Methods). Figure~\ref{fig:fig4}(b) shows that they do not. Instead, we find that resolving local rather than global phonon characteristics minimizes memory times everywhere (SI~Fig.~S4), identifying locally resolved bath distortions as the essential slow variable. Since the kernel at the seam remains as long-lived as those on either side, the seam carries memory but produces no \textit{net} delay to thermalization.

To understand how global versus locally resolved lattice relaxation control memory, we resolve distance-dependent contributions to $\tau_M$ using the kernel extracted from tensor-network dynamics (Methods). Weighting each transfer by its squared distance, we decompose $\tau_M = \sum_{|R|\geq 1} \tau_{M,|R|}$ into contributions $\tau_{M,|R|}$ from transfer across $|R|$ sites (Methods), each a head start ($\tau_{M,|R|}<0$) or a delay ($\tau_{M,|R|}>0$). Figure~\ref{fig:fig4}(c) shows this decomposition. In the subdiffusive case, $\tau_M \approx -12$~fs, and nearest- and next-nearest-neighbor transfer accounts for nearly all of it, recapitulating the physical picture of local Marcus-type hopping. The negative memory is local, confined to the site where the polaron forms. In the superdiffusive case, $\tau_M \approx +27$~fs has contributions from many sites, with the first providing only $\approx3$~fs. The dressed carrier thus moves coherently across several sites, and this long-range transfer, which deviates from the Markovian Marcus picture, delays the onset of diffusion. While explicitly connecting only nearest-neighbor coupling, the master equation behind Fig.~\ref{fig:fig3} (Methods) is a non-Markovian generalization of Marcus theory and qualitatively recovers the sign of $\tau_M$ and the seam. The diffusive seam is where the local head start and the non-local delay cancel.

\section*{Outlook}

Employing the dispersive Holstein model, we show that where the phonon bath sits relative to the electronic band sets the transient transport regime of a relaxing carrier. Resonant baths hold the carrier in subdiffusion, off-resonant baths drive it into superdiffusion, and a narrow seam between them yields diffusion swiftly after injection. This control is within experimental reach, since all three controls are synthetically tunable---the mode frequency through chemical or isotopic substitution, the mode lifetime through anharmonicity, and the initial condition through the excitation energy---and the resulting $\mathrm{MSD}(t)$ is directly resolvable by spatiotemporal microscopies~\cite{penwell2017resolving, ginsberg2020spatially}. Because transient subdiffusion arises from polaron formation, it should vanish for off-resonant baths, for carriers injected near their polarized geometry, and in measurements of terahertz conductivity (Methods).

More broadly, out-of-equilibrium many-body physics, from Griffiths bottlenecks to prethermal plateaus~\cite{Agarwal2015, Gring2012}, has largely lacked a unifying quantitative metric of how much a system transiently departs from its hydrodynamic behavior. Since these departures live in the low-frequency structure of the memory kernel, $\tau_M$, its higher-order counterparts, and $N_c$ can quantify how far, in what direction, and for how long a system is held from equilibrating, and outline a path toward systematic control of thermalization in nonequilibrium many-body systems. 

\bibliography{main.bib}

\section*{Methods}

\MethodsCitations

\subsection*{Transport observables}

We compute the MSD on the periodic $N$-site ring from the site populations, $\mathrm{MSD}(t) = a^2\sum_n d_n^2 P_n(t)$, where $d_n = \min(|n-n_0|, N-|n-n_0|)$ is the distance from the initially occupied site $n_0$ defined under periodic boundary conditions. The transport exponent is $\alpha(t) = d\ln\mathrm{MSD}/d\ln t$.

Once population wraps around the ring, the MSD saturates and $\alpha(t)$ falls towards zero. We therefore define the onset time of finite-size effects, $t^*$, as the first time the site farthest from $n_0$ acquires a threshold population $\epsilon$, and set $\alpha^* \equiv \alpha(t^*)$. The choice of $\epsilon$ balances two errors. A small $\epsilon$ sets $\alpha^*$ early, while the carrier is still relaxing, and hence overstates $|1-\alpha^*|$. A larger $\epsilon$ sets $\alpha^*$ after population reaches the boundary, reaching finite size effects. We use $\epsilon = 10^{-4}$. We extract $N_c$ by a least-squares fit of $\ln|1-\alpha^*| = -N/N_c$ over $N=10-50$. This fit is descriptive, and it summarizes how far the departure from diffusion persists over the lattice sizes we simulate. We rely on $\tau_M$ to faithfully describe the asymptotic approach to diffusion directly from the memory kernel, without reference to lattice size or to the threshold $\epsilon$. Near the seam, $|1-\alpha^*|$ is near zero at every $N$ and the fit is ill-conditioned (see Fig.~\ref{fig:fig2}(b));  hence, we interpolate the transport diagram in Fig.~2 on the $(\Omega,\gamma)$ grid around that region.

\subsection*{Population GME and NIBA kernel}

We obtain Eq.~\ref{eq:gme} by projecting the full dynamics onto the site populations $\lvert i\rangle \langle i\rvert$ ($\hbar = 1$ throughout). Because the exact kernel is intractable across our parameter range, we adopt the non-interacting blips approximation (NIBA), which treats the hopping $v$ to second order and the carrier-phonon coupling to all orders~\cite{dekker1987noninteracting}, so that transfer occurs only between nearest neighbors. The bath enters only through the functions given by

\begin{align}
  Q_1(t) &= \frac{2}{\pi}\int_0^\infty d\omega\, \frac{J(\omega)}{\omega^2}\coth\left(\frac{\beta\omega}{2}\right)\left[1-\cos(\omega t)\right], \label{eq:Q1}\\
  Q_2(t) &= \frac{2}{\pi}\int_0^\infty d\omega\, \frac{J(\omega)}{\omega^2}\sin(\omega t). \label{eq:Q2}
\end{align}

The kernel depends on how the lattice is prepared at $t=0$. Our default is a Franck-Condon (FC) excitation, in which the carrier is placed on a lattice equilibrated in its absence. The resulting kernel is non-stationary:
\begin{equation}
  \mathcal{K}_{\mathrm{FC}}(t,s) = 2|v|^2 e^{-Q_1(t-s)} \cos\left[Q_2(t-s) - Q_2(t) + Q_2(s)\right].
  \label{eq:kernel-FC}
\end{equation}

whereas for a lattice polarized around the carrier at $t=0$ the phase $Q_2(t)-Q_2(s)$ vanishes~\cite{mendes2026coherent}, resulting in a stationary kernel
\begin{equation}
  \mathcal{K}_{\mathrm{rel}}(t-s) = 2|v|^2 e^{-Q_1(t-s)} \cos\left[Q_2(t-s)\right].
  \label{eq:kernel-rel}
\end{equation}

Unless stated otherwise, $\lambda = 2700~\unit{cm}^{-1}$ and $k_BT = 695~\unit{cm}^{-1}$, for which NIBA reproduces numerically exact HEOM~\cite{tanimura89, tanimura20} up to the onset of finite-size effects (SI~Sec.~I).

\subsection*{Memory Kernel Theory of Transport}

\textit{Bath memory time} --- For a homogeneous lattice with lattice constant $a$ and the carrier initialized on site $0$, the memory kernel for the population GME depends on the separation $R = n-m$, $\mathcal{K}_{nm}(t,s) = \mathcal{K}_R(t,s)$. From population conservation, $\sum_R \mathcal{K}_R = 0$, and by symmetry, we get $\mathcal{K}_R = \mathcal{K}_{-R}$. Inserting Eq.~\ref{eq:gme} into $\mathrm{MSD}(t) = a^2\sum_n n^2 P_n(t)$ and using these two properties gives

\begin{equation}
\begin{split}
  \frac{d\mathrm{MSD}}{dt} & = 2D(t) = \int_0^t ds\ \tilde{\mathcal{K}}(t,s), \\
  \tilde{\mathcal{K}}(t,s) &= a^2 \sum_R R^2 \mathcal{K}_R(t,s).
  \label{eq:Dt}
\end{split}
\end{equation}

Once the kernel has decayed, $D(t) \to D_\infty$ and $\mathrm{MSD}(t) \to 2D_\infty(t-\tau_M)$, with

\begin{equation}
  \tau_M = \int_0^\infty dt \left[1 - \frac{D(t)}{D_\infty}\right].
  \label{eq:tauM-general}
\end{equation}

For a one-time kernel, Eq.~\ref{eq:tauM-general} reduces to the normalized first moment form provided in the main text, $\tau_M = -\hat{\tilde{\mathcal{K}}}'(0)/\hat{\tilde{\mathcal{K}}}(0)$. For a Markovian kernel, $\tilde{\mathcal{K}}(t,s)\propto\delta(t-s)$, it becomes zero. Differentiating the asymptotic MSD gives $\alpha(t) = t/(t-\tau_M) \simeq 1 + \tau_M/t$ for $t \gg |\tau_M|$, so that $\mathrm{sgn}(1-\alpha^*) = -\mathrm{sgn}(\tau_M)$.

\textit{Terahertz conductivity} --- For an equilibrium initial state, $d^2\mathrm{MSD}/dt^2 = 2C_v(t)$, where $C_v(t)$ is the velocity autocorrelation function, so that $\tilde{\mathcal{K}}(t) = 2C_v(t)$. Linear response theory then gives the frequency-dependent conductivity as $\sigma(\omega) \propto \hat{\tilde{\mathcal{K}}}(-i\omega)$. Expanding $\hat{\tilde{\mathcal{K}}}(s)$ to first order in $s$, we get $\sigma(\omega) \simeq \sigma_{\rm DC}(1 + i\omega\tau_M)$, i.e., $\mathrm{Im}\ \sigma(\omega) \simeq \sigma_{\rm DC}\ \omega\tau_M$ at low frequency. The terahertz conductivity approximately measures $\tau_M$ of the kernel beginning from a polarized initial condition (Eq.~\ref{eq:kernel-rel}), which lacks the negative memory of polaron formation (Fig.~\ref{fig:fig4}(a)).

\textit{Changing the projection operator} --- We built the population-based GME by projecting onto the site populations ($\lvert i \rangle \langle i \rvert$) of the Holstein model. Projecting instead onto an arbitrary set of variables $\{A_i\}$, $\mathcal{P} = |\mathbf{A})(\mathbf{A}|$, yields a Volterra equation for the correlation matrix,

\begin{equation}
    \dot{\mathcal{C}}(t) = \mathcal{C}(t)\dot{\mathcal{C}}(0) -\int_0^{t}d\tau \mathcal{C}(t-\tau)\mathcal{K}(\tau).
    \label{eq:corr-gme}
\end{equation}

which we invert to obtain the memory kernel. We generate $\mathcal{C}(t)$ for arbitrary projections from exact quantum dynamics (such as the one we describe in the Tensor Network Method section below). To include electronic coherences~\cite{Argyres1964}, we project onto $|j\rangle\langle k|$, with $\mathcal{C}_{jk,lm}(t) = \rho_{jk}(t)$ for $\rho(0) = |l\rangle\langle m|$. To include local lattice distortions, we project onto $A_{i,a} = \hat{n}_i\hat{O}_{a,i}$ with $\hat{O}_a \in \{\hat{1}, \hat{X}, \hat{P}\}$, where $\hat{X}_i = \hat{b}_i + \hat{b}_i^{\dagger}$ and $\hat{P}_i = i(\hat{b}_i^{\dagger} - \hat{b}_i)$. Then $\mathcal{C}_{ia,jb}(t) = G_a^{-1}\mathrm{Tr}[\hat{n}_i\hat{O}_{a,i}\rho_{j,b}(t)]$ with $\rho_{j,b}(0) = |j\rangle\langle j| \otimes \hat{M}_{b,j} \otimes_{k\neq j} \rho_{\mathrm{th},k}$, where $\hat{M}_b \in \{\rho_{\mathrm{th}}, \tfrac{1}{2}\{\hat{X},\rho_{\mathrm{th}}\}, \tfrac{1}{2}\{\hat{P},\rho_{\mathrm{th}}\}\}$ and $G_1 = 1$, $G_X = G_P = 2\bar{n}+1$ ensure $\mathcal{C}(0) = \mathbb{1}$.. We define the kernel lifetime $\tau_c$ as the shortest truncation time for which setting $\mathcal{K}(t>\tau_c) = 0$ in Eq.~\ref{eq:corr-gme} reproduces the exact MSD to a normalized root mean-squared error below $10^{-3}$ (see SI Sec.~III).

\textit{Decomposition by distance} --- For a one-time kernel, $\tilde{\mathcal{K}}(t) = a^2\sum_{|R|\geq1} 2R^2\mathcal{K}_R(t)$, so its first moment splits exactly into contributions from transfer over each distance as $\tau_M = \sum_{|R|\geq 1} \tau_{M,|R|}$, where
\begin{equation}
  \tau_{M,|R|} = \frac{2a^2R^2\int_0^\infty dt [t\mathcal{K}_R(t)]} {\int_0^\infty dt \tilde{\mathcal{K}}(t)}.
  \label{eq:tauM-shell}
\end{equation}
Each term is normalized by the total kernel weight so the contributions add to $\tau_M$. In this work, we split the kernel generated from projections onto site populations as well as phonon displacements and momenta into distance-dependent components.

\subsection*{Tensor Network Method}

To project onto degrees of freedom that include phonon coordinates, we represent each Lorentzian bath by a discrete phonon mode of frequency $\Omega$ and coupling $g = \sqrt{\lambda\Omega}$. To shorten the lifetime of the mode, we attenuate it by a Lindbladian with damping $2\gamma$. The density matrix of the carrier and the modes obeys~\cite{garraway1997}:
\begin{equation}
  \frac{d\rho}{dt} = -i[H,\rho] + 2\gamma\sum_i \Big((\bar{n}+1)\mathcal{D}[\hat{b}_i]\rho + \bar{n}\mathcal{D}[\hat{b}_i^{\dagger}]\rho \Big),
  \label{eq:lindblad}
\end{equation}
where $H$ is Eq.~\ref{eq:holstein-hamiltonian} with one mode per site, $\mathcal{D}[\hat{x}]\rho = \hat{x}\rho\hat{x}^{\dagger} - \frac{1}{2}\{\hat{x}^{\dagger}\hat{x},\rho\}$, and $\bar{n} = (e^{\beta\Omega}-1)^{-1}$ so that each mode relaxes to its thermal state.

We write Eq.~\ref{eq:lindblad} as $d\lvert\rho\rangle\rangle/dt = \mathcal{L}\lvert\rho\rangle\rangle$ for the vectorized density matrix and represent $\lvert\rho\rangle\rangle$ as a matrix product state (MPS) on a ring of $N = 15$ sites. We truncate phonon modes to $d = 20$ Fock states and propagate real-time dynamics with the time-dependent variational principle~\cite{Haegeman2016} using a matrix product operator representing the Liouvillian $\mathcal{L}$. We show in SI~Sec.~I that NIBA under the Brownian oscillator spectral density qualitatively agrees with the dynamics produced with the damped mode. We use NIBA to map $\tau_M$ across the $(\Omega,\gamma)$ plane (Figs.~\ref{fig:fig1}--\ref{fig:fig3}) and tensor networks to resolve its mechanism at three representative baths (Fig.~\ref{fig:fig4}(b),(c)).

\methodsbibliography 
\end{document}